\documentclass{aa}  
\usepackage[dvips]{graphicx}
\usepackage{txfonts}

\begin{document}

\newcommand{\fe}{\ion{Fe}{II}}
\newcommand{\h}{H$_2$}
\newcommand{\kms}{km\,s$^{-1}$}
\newcommand{\cmt}{cm$^{3}$}
\newcommand{\cmd}{cm$^{-2}$}

\title{Warm SiO gas in molecular bullets associated with protostellar
outflows}
\author{B. Nisini\inst{1}, C. Codella\inst{2}, T. Giannini\inst{1}, 
J. Santiago Garcia\inst{3}, J.S. Richer\inst{4}, R. Bachiller\inst{3}, M. Tafalla\inst{3}}
\offprints{Brunella Nisini, nisini@oa-roma.inaf.it}

\institute{INAF-Osservatorio Astronomico di Roma, Via di Frascati 33, 
I-00040 Monteporzio Catone, Italy  \and INAF-Istituto di Radioastronomia, Sezione di Firenze, 
Largo E. Fermi 5, I-50125 Firenze, Italy \and 
Observatorio Astron\'omico Nacional (IGN), Alfonso XII 3, 28014 Madrid, Spain
\and Cavendish Laboratory, J J Thomson Avenue, 
Cambridge CB3 0HE, UK}
%
%
\date{Received date; Accepted date}
%
%
%
\titlerunning{Warm SiO in molecular bullets}
\authorrunning{B. Nisini et al.}

 
  \abstract
   {In this paper we present the first SiO multiline analysis (from $J$=2-1 to $J$=11-10) 
   of the molecular
bullets along the outflows of the Class 0 sources L1448-mm and L1157-mm, obtained through
observations with IRAM and JCMT.}
   {This analysis has been performed to investigate the physical properties of  
   different types of bullets, in order to explore their excitation conditions and put constraints on
   their formation and evolution.}
   {For this purpose we have computed the main physical parameters ($n_{H_2}$, $T_{kin}$, $N_{SiO}$) in each bullet
   and compared them with other tracers of warm and dense gas and with models for the SiO excitation in shocks.}
   {We find that the bullets close to L1448--mm, associated with high velocity gas, have higher
   excitation conditions ($n_{H_2} \sim$10$^{6}$ cm$^{-3}$, $T \ga 500$ K) with respect to the L1157 bullets
   ( $n_{H_2} \sim$1--5$\times 10^{5}$ cm$^{-3}$, $T \sim 100-300$ K).  In both the sources, there is a clear evidence of the presence of velocity components
   having different excitation conditions, with the denser and/or warmer gas associated with 
the gas at the higher speed. 
In L1448 the bulk of the emission is due to the high-excitation and high velocity gas, while in L1157 most of the emission comes from the low excitation gas at ambient velocity.
The observed velocity-averaged line ratios 
are well reproduced by shocks with speeds v$_{s}$ larger than 
$\sim$30 \kms\, and densities $\sim$ 10$^{5}$ --
  10$^{6}\,$cm$^{-3}$. Plane-parallel shock 
models, however, fail to predict all the observed line profiles and in particular the very similar profiles shown by both low and high excitation lines. }
   {The overall observations support the idea that the L1157 clumps 
   are shock interaction events older than the L1448 bullets 
   close to the driving source. In the latter objects, the velocity structure 
   and the variations of physical parameters with the velocity 
   resemble very closely those found in optical/IR jets near the 
   protostar, suggesting that similar launching and excitation 
   mechanisms are also at the origin of collimated jets seen at
  millimetre wavelengths.  
}

\maketitle
%
%


\section{Introduction}

The mass-loss process starts during the earliest phases of star
formation when protostars are still accreting mass (e.g. Richer et al. \cite{richer},
and reference therein). The stellar winds create jets that interact with the high-density 
clump hosting the newly-born stars, forming shocks and driving molecular outflows.
Fast ($\ge$ 100 km s$^{-1}$) collimated jets are observed both in optical emission of
atomic lines and near infrared (NIR) H$_2$ emission, showing excitation conditions
with $T_{\rm ex}$ ranging from $\simeq$ 2000 K to more than 10\,000 K. 
Also, extremely high velocity (EHV) jets have been seen in the
form of narrow and high velocity CO and SiO molecular outflows in a handful of Class 0 sources
(e.g. L1448-mm, HH211, IRAS04166, Gu\'eth \&  Guilloteau \cite{guethgui},
Tafalla et al. \cite{tafalla}).
It is a matter of some debate whether the EHV molecular jet is the cold  external layer of an embedded atomic
jet, or whether the jet itself is intrinsically molecular. 
In particular, the EHV molecular jets show a knotty structure with regularly spaced
clumps, the so-called molecular bullets, that are well-defined entities travelling at
velocities of hundreds of km s$^{-1}$ (e.g. Bachiller et al. \cite{bach90}, Chandler \& Richer
\cite{chandler}). The bullets are most likely associated with shocks formed by
the outflow propagation into a homogeneous material and are consequently thought
to be a direct manifestations of the episodic nature of outflows (Dutrey et al. \cite{dutrey},
Arce \& Goodman \cite{arce1,arce2}).
On the other hand, a few outflows driven by Class 0 objects (e.g. L1157-mm, NGC2264G, BHR71)
show another type of molecular clump, 
 characterised by lower velocities and by a rich chemistry:
these clumps have strong  
emission due to molecular species which increase their abundance only
when a shock-induced chemistry is active (Bourke et al. \cite{bourke}, Bachiller et al.
\cite{bach01}). The origin of these chemically rich clumps is not clear:
a possibility is that they are a 
consequence of a pre-existing density enhancement and of its interaction with
the passage of a shock (Viti et al. \cite{viti}).

However, the nature of both kinds of clumps and the possible relationships between them
based on age, physical characteristics, and/or excitation conditions, are not yet understood.
Detailed quantitative analysis of the EHV jets, bullets, and chemically rich clumps have
been so far obtained with two standard outflow tracers such as CO and SiO, but  
only with low frequency and consequently low excitation transitions. These transitions,
 however, are not suited to probe the warm gas components that may be
associated with high-velocity and shocked gas. Indeed, sub-mm-wavelengths observations 
show that some molecular bullets are dense, $n_{H_2}$ $\sim$ 10$^6$ cm$^{-3}$, and
warm, $T_{\rm ex}$ $\ge$ 100 K (Hatchell et al. \cite{hatchell}, Nisini et al. \cite{nisini02},
Palau et al. \cite{palau}).

In addition, ISO-LWS high-J CO and H$_2$O low angular resolution observations 
seem to suggest that 
EHV jets are associated with temperatures in the 300--1500 K range (Giannini et al. \cite{gia}).
Whether this warm gas represents a major component of the mass flux and consequently of the
energetics of the mass loss from a YSO is still an open question.

In this paper, we investigate the nature of both the high-velocity bullets and
chemically-rich clumps through the use of a multi-frequency analysis of the SiO emission. 
This molecule is the most typical tracer of high-density conditions and high 
velocity shocks, strong enough to release refractory elements in the gas phase 
(e.g. van Dishoeck \& Blake \cite{van}).
So far, SiO has been mainly observed at high velocities through its low rotational transitions
($J$ $\le$ 5), tracing relatively cold ($\le$ 100 K) gas 
(Mart\'{\i}n-Pintado et al. \cite{martin}, Codella et al. \cite{cod99}, Gibb et al. \cite{gibb}).

 The objects studied in this paper are the outflows L1448 and L1157,
representative of flows associated with EHV jet-like bullets
and lower velocity clumps respectively. 
In particular, here we complement the low excitation spectra of SiO $J$=2--1, 3--2 and 5--4,
with new observations of the $J$ = 6--5, 8--7, and 11--10 emission at submillimetre wavelengths.
High-$J$ SiO emission from $J$ = 5--4 to 11--10 has been already observed in 
another typical EHV flow, HH211, by Nisini et al. (2002). Here we perform 
a detailed analysis of the full set of lines with the following aims:  
(i) disentangle, through their excitation conditions, the warm gas components directly
originating in the EHV jet from ambient and colder gas entrained by the jet itself;  
(ii) compare the spectra due to different
transitions and thus investigate whether the excitation conditions vary with velocity, 
comparing the results with the predictions given by shock models; and (iii) estimate 
the type and strength of the shocks, and so place constraints on the origin 
of bullets and chemically rich clumps.

\section{Observations and results}

Observations were carried out with the IRAM 30-m telescope for the SiO
transitions from $J$ = 2--1 to 6--5, and with the  
James Clerk Maxwell Telescope (JCMT) for the SiO transitions from $J$ = 8--7 
to 11--10. In Table 1, a list of the lines observed and their excitation
energy is presented together with relevant observational parameters.

JCMT observations were performed during several nights in November 2001, 
and October and November 2002. We used the heterodyne receiver 
RxW for the $J=11-10$ and $J=10-9$ 
observations (HPBW = 12\arcsec), and RxB3 for the $J=8-7$ observations  
at 347\,GHz (HPBW = 14\arcsec). 
The back-end was the digital autocorrelation spectrometer (DAS) with
a bandwidth of 500 MHz. 
To convert the antenna temperature ($T_{\rm A}^*$) into main-beam brightness temperatures, 
main-beam efficiencies $\eta_{MB}$ equal to 0.55 and 0.4 were adopted for 
receivers B3 and W respectively. Pointing was checked on NGC7538 IRS1 and W3(OH) 
and was good within about 3\arcsec.

\begin{figure}
\centerline{\includegraphics[angle=0,scale=0.2]{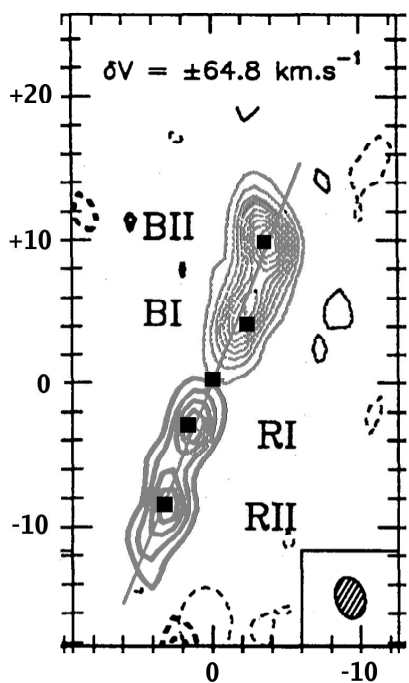}}
\caption{\label{fig:1448_map} The pointings adopted for our observations of the central region of L1448 are
indicated by black squares over an interferometric velocity channel map at 64.8 \kms\,of the SiO 2--1 transition taken from 
Guilloteau et al. (1997). Offsets (in arcsec) are with respect to L1448-mm.}

\label{fig1}
\end{figure}

In L1448, we observed strips of 5 and 3 points in the SiO 8--7 and 11--10 lines 
respectively, along the 
major axis of the outflow (P.A. 157$\fdg$4) with a 10$\arcsec$ spacing (see Fig.\ref{fig:1448_map}). 
The (0,0) position of this strip corresponds to
the exciting source of the outflow, L1448-mm ($\alpha_{2000}$=03$^{h}$22$^{m}$34.3$^{s}$,
$\delta_{2000}$=30$^{d}$33$^{m}$35$^{s}$).

 The pointings along the strip covers the CO bullets R1 and R2 identified by Bachiller et al. (1990),
resolved by the Guilloteau et al. (1992) interferometric observations in two distinct peaks 
each, named BI--BII and RI--RII and indicated in Fig.\ref{fig:1448_map}.

In the (0,0) position a spectrum of the SiO 10--9 transition was also obtained.
Spectra of the pointings centered on the mm source
and on the B1,R1 bullets are shown in Fig. \ref{fig:1448-spec}.
In addition to the central region, a SiO 8--7 spectrum was obtained also in the south lobe 
of the outflow, at the position were a strong SiO 2--1 peak was observed 
by Dutrey et al. 1997 (position R4 following Bachiller et al. (1990)). 
\\

\begin{figure}
\centerline{\includegraphics[angle=-90,scale=0.8]{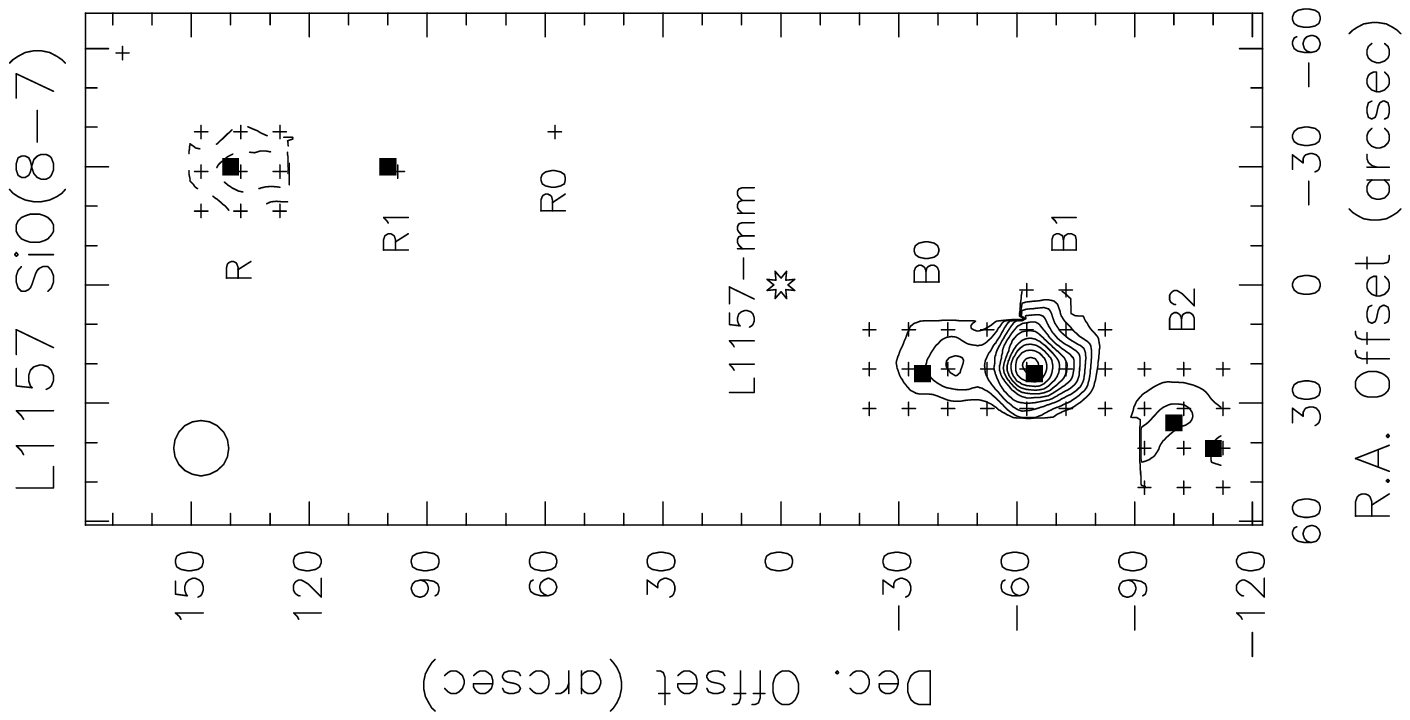}}
\caption{\label{fig:1157map}Map of the SiO $J$ = 8--7 integrated emission in the L1157 outflow. 
Offsets are with respect to L1157-mm. First contour and steps are 0.75 K \kms\,and 0.60 K \kms\,for 
the blue and red bullets, respectively (which correspond to 3$\sigma$ in each case). 
The circle represents the JCMT HPBW, while the small crosses indicate the observed positions.
 Filled squares indicate the pointings of the IRAM data considered in our analysis.}
\label{fig1}
\end{figure}

\begin{figure}[!ht]
\resizebox{\hsize}{!}{\includegraphics[scale=0.5]{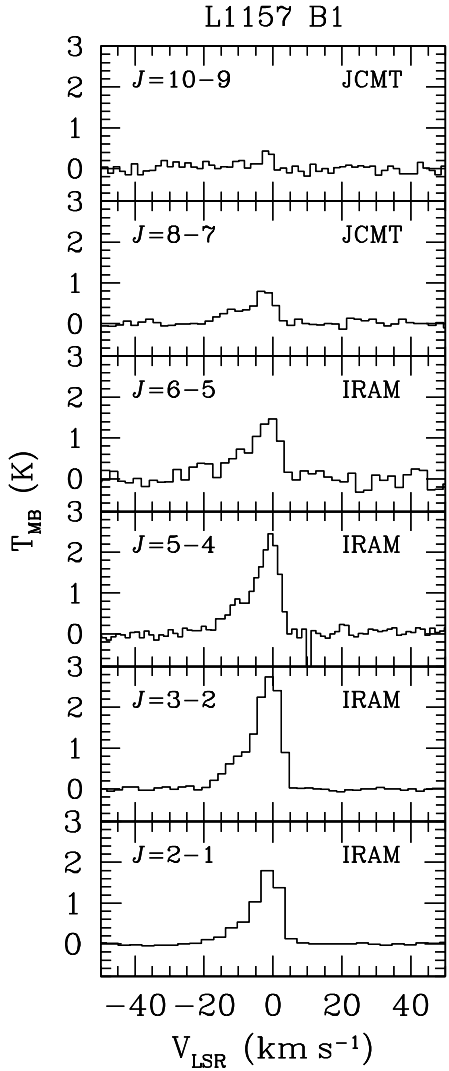}}
\caption{\label{fig:1157-spec}Observed SiO spectra towards the L1157 B1 position. }
\end{figure}

\begin{figure*}[!ht]
\resizebox{\hsize}{!}{\rotatebox{0}{\includegraphics{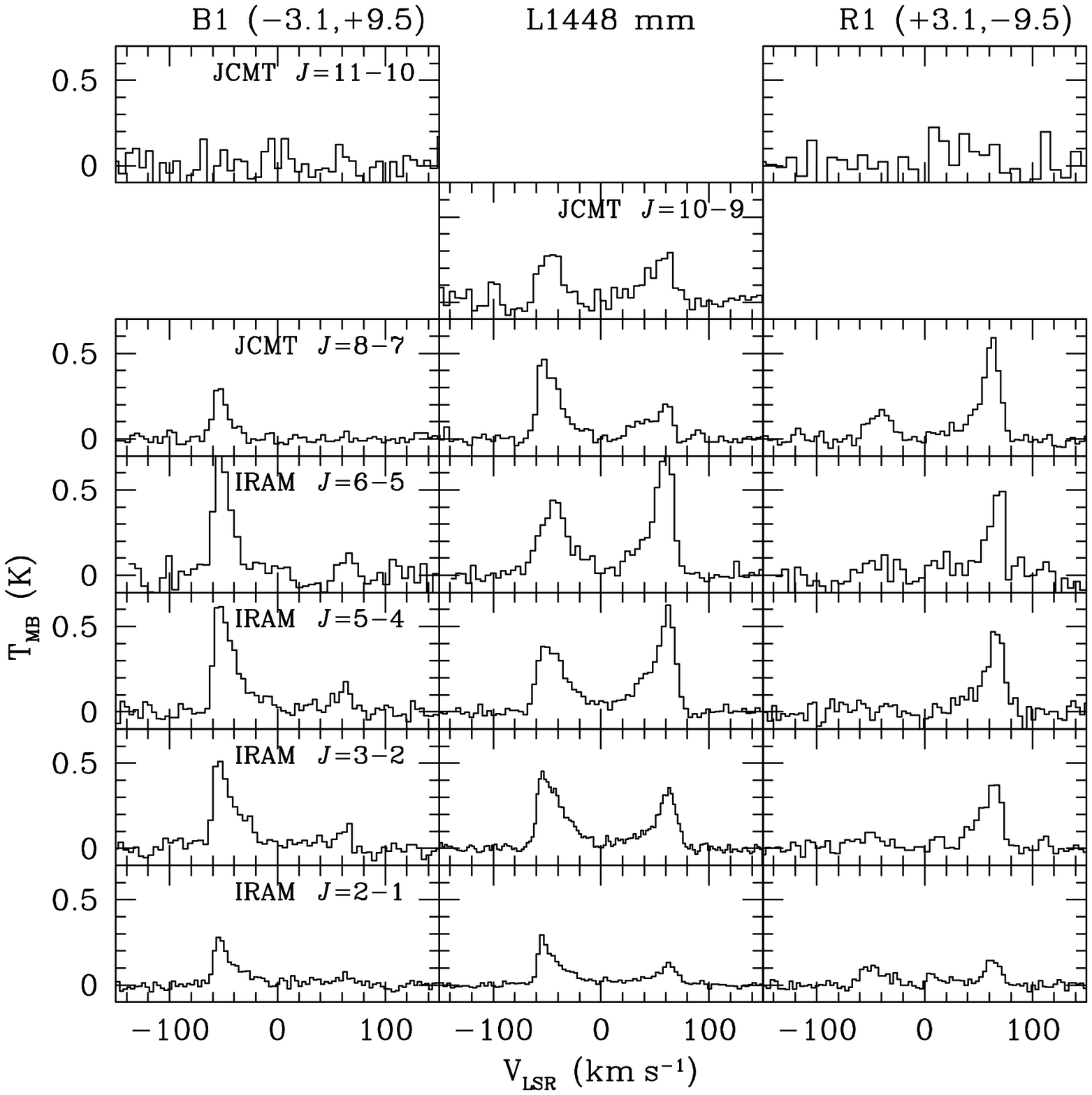}}}
\caption{\label{fig:1448-spec}Spectra of the SiO line emission taken in positions at the base of the
L1448 outflow. Offsets (in arsec) of the B1 and R1 positions are with respect to the L1448-mm
coordinates.}
\end{figure*}

In L1157, a map in the SiO 8--7 line was acquired covering the strongest bullets
identified by Bachiller et al.\cite{bach01} , B0, B1 and B2 in the southern lobe
of the outflow, and R0, R1 and R in the northern lobe (Fig. 
\ref{fig:1157map}).
Additional spectra of the SiO 10--9 transitions were obtained only in the B0, B1
and B2 positions. Fig. \ref{fig:1157-spec} shows the spectra of the brightest
B1 peak.\\

The JCMT observations were complemented by observations of SiO transitions
at lower frequency obtained at nearby positions using the IRAM-30m telescope 
(Santiago Garcia, PhD thesis). 
These observations were performed on the 14th June 2000. 
Heterodyne receivers A100, A230, D150, and D270 were tuned to observe all the lines simultaneously 
in wobbler-switching mode, with HPBW beams from 28\arcsec to 9\arcsec, 
and a wobbler throw of $\sim$240\arcsec. 
For all the transitions an integration time of 4 minutes was accumulated in all the 
L1157 bullet spectra, and in the B1, R1 and R4 spectra in L1448, 
while the spectrum of the central L1448-mm position was integrated for 8 minutes. 
Pointing was checked in nearby sources and found to be accurate within 2\arcsec. 
System temperatures ranged from $\approx$166 K -- 208 K for the J=2--1 line, to $\approx$ 1016 K -- 2180 K  for the J=6--5 line. 
Spectra obtained with IRAM are also shown in Figs. \ref{fig:1157-spec} and \ref{fig:1448-spec},
together with the JCMT spectra.
\\

For both JCMT and IRAM observations, we have obtained the 
intensity integrated over the velocity channels, separating the blue-
and red-shifted components.  These values are listed in Tables 2 and 3 
only for the positions which will be relevant for our analysis. 
The errors on the integrated intensities reported in these tables are the 
statistical errors due to thermal noise on the
individual spectra.  Calibration accuracy is about 10\% for lines with $J\le$8, and 20\% for the 
$J$=10--9 and 11--10 lines, and these uncertainties must be considered when estimating the total errors 
on observed quantities.
In the L1157 flow, the coordinates of the beam centers in the JCMT observations do 
not always coincide with those
of the IRAM observations, in which case we report in Table 3 the intensities 
obtained from the spectra closer in space. The angular difference of these spectra is 
typically of the order of 2--3\arcsec: in Fig. \ref{fig:1157map}, the different displacement of
 the IRAM and JCMT pointings is indicated. 
\\

Comparing the L1157 SiO $J$= 8--7 map with the SiO $J$= 3--2 and 5--4 maps 
in Bachiller et al. (2001), we see a similar brightness distribution: 
the strongest SiO emission is associated with the blue lobe, and in particular with
bullet B1, the one most chemically active along the flow. In our map, we resolve 
the peak of the B0 bullet that remained unresolved in the 5--4 IRAM maps.
We do not detect any SiO 8--7 emission in the R1 and R0 bullets.
We also find a general agreement between the 
high-J SiO distribution and the distribution of the warm gas mapped 
through the pure rotational H$_2$ 0--0 transitions by ISOCAM (Cabrit et al. 1999).

In L1448 the situation is somehow different. The strongest high-J 
lines are detected close to the millimetre source, in the B1 and R1 bullets, 
while the R4 bullet in the southern lobe, although very strong in 
SiO 2--1, is much weaker in the high-$J$ lines. This behaviour was already 
pointed out by Bachiller et al. (1991) but this result is much clearer 
in our 8--7, 10--9 and 11--10 observations. 
In L1448 there is no H$_2$ 0--0 map to compare the morphology of 
the SiO emission with the occurrence of warm gas; however, we point out that the R1 and 
B1 bullets are those 
associated with the region of the strongest high-J CO and H$_{2}$O emission 
observed by ISO-LWS (Nisini et al. 2000). 

The SiO 2--1 interferometric observations of Guilloteau et al. 
(1992) have shown that there is a velocity pattern along
the SiO jet, with the velocity linearly increasing from BI--RI (at 3--4\arcsec) to BI--RII (at $\sim$10\arcsec)
. We 
have evidence of this behaviour in the shapes of the lines at 
different frequency, which are observed with different beam sizes. In the observations centered on the mm source, 
the blue and red-shifted emission
peaks are located at velocities which decrease going from the 2-1 (V$_{LSR}$ = $-$55 and +63\kms\ for the B1 and R1 bullets respectively) to the 6--5 lines 
(V$_{LSR}$ = $-$43.2 and +68.1\kms), this latter having a beam size of only 9\arcsec\ and 
thus including only the slower BI--RI components. 

\begin{table*}
\caption{Observed SiO transitions and instrumental setting}
\begin{tabular}{lccccccc}
\hline
Line & $E_{up}$ & Instr. & Freq. & Res. & HPBW & $\eta_{B}$ 
\\
 & K& &  GHz & km\,s$^{-1}$ & $^{\arcsec}$ &\\
\hline
SiO $J$=2--1  & 6.2  &IRAM         & 86.85   & 3.45 & 27.6 & 0.77 \\
SiO $J$=3--2   & 12.5 &IRAM         & 130.27  & 2.30 & 18.4 & 0.64 \\
SiO $J$=5--4  & 31.2 &IRAM         & 217.10  & 1.38 & 11.0 & 0.56 \\
SiO $J$=6--5   & 43.7&IRAM         & 260.52  & 1.15 & 9.2 &  0.51 \\
SiO $J$=8--7  & 75.0 &JCMT     & 347.33  & 1.10 & 14 & 0.55 \\
SiO $J$=10--9 & 114.5 &JCMT     & 434.12  & 1.73 & 11 & 0.4  \\
SiO $J$=11--10 & 137.4&JCMT     & 477.50  & 0.79 & 11 & 0.4 \\
\hline	    

\end{tabular}
\end{table*}

\begin{table*}
\caption {Observed SiO intensities in L1448-mm}
\vspace{0.5cm}
\begin{tabular}[t]{@{}ccccccccc}
\hline
 Offset$^{a}$ &  Bullet & $J$=2--1 & $J$=3--2 & $J$=5--4 & $J$=6--5 &  $J$=8--7 &  $J$=10--9 &  $J$=11--10 \\
  &    & \multicolumn{7}{c}{$\int T_{MB}\,dV^{b}$} \\
\hline
 ($-$3.1, +9.5)&   B1 & 6.3$\pm$0.3 & 14.0$\pm$0.4 & 15.8$\pm$0.7 & 16.7$\pm$1.2 & 5.3$\pm$0.5 &  ... & $<$1.2\\
             &  R1  & 1.9$\pm$0.3 & 3.2$\pm$0.4 & 3.8$\pm$0.7 &... & ...& ...& $<$1.2\\
\hline  
 ($-$2.3, +3.4) &   B1 & 7.0$\pm$0.4 & 13.0$\pm$0.4 &15.8$\pm$0.7 & 19.0$\pm$1.2 & ... &... &...\\
              &   R1  & 3.7$\pm$0.4 & 7.7$\pm$0.4 & 11.2$\pm$0.7 & 12.7$\pm$1.2 & ... &... &...\\
\hline
 (0,0)        &   B1 & 6.6$\pm$0.2 & 12.5$\pm$0.4 & 13.0$\pm$0.4 & 14.6$\pm$0.6 & 12.0$\pm$0.5 & 6.7$\pm$0.9 &$<$1.2\\ 
              &   R1  & 3.8$\pm$0.2 & 9.4$\pm$0.4 & 16.5$\pm$0.4 & 18.8$\pm$0.6 & 11.6$\pm$0.5 & 8.7$\pm$0.9 &4.5$\pm$0.9\\
\hline
 (+2.3, $-$3.4) &   B1 &5.1$\pm$0.4 & 6.8$\pm$0.4 & 6.4$\pm$0.7 & 3.5$\pm$1.1& ... & ... &...\\
              &   R1  & 4.4$\pm$0.4 & 9.4$\pm$0.4   & 13.5$\pm$0.7 &  20.6$\pm$1.1 & ... &... & ...\\
\hline
 (+3.1, $-$9.5) &   B1 & 2.6$\pm$0.4 & 2.6$\pm$0.5 & ... & 1.4$\pm$1.2 & 4.0$\pm$0.4 &... &$<$1.2\\
              &   R1  & 4.0$\pm$0.4 &10.0$\pm$0.5  & 11.1$\pm$0.7 &  11.5$\pm$1.2 & 13.3$\pm$0.4& ... & 6.7$\pm$0.9\\  
\hline
 (+26,$-$128)   &   R4 & 13.1$\pm$0.3 & 17.1$\pm$0.5 & 10.7$\pm$0.5 & 8.3$\pm$1.2 & 3.1$\pm$0.4 & &\\	    
\hline
\end{tabular}

$^a$ Offset in arcsec with respect to L1448-mm ($\alpha_{2000}$=03$^{h}$22$^{m}$34.3$^{s}$, 
$\delta_{2000}$=+30$^{d}$33$^{m}$35$^{s}$)\\
 $^b$ Flux errors refer to the rms noise in the individual spetra only. Total calibration uncertainty are of the order of 10\% for the $J\le$8
lines and 20\% for the $J$=10 and 11 lines.
\end{table*}

\begin{table*}
\caption{Observed SiO intensities in L1157}
\vspace{0.5cm}
\begin{tabular}[t]{@{}cccccccc}
\hline
 Offset$^{a}$ &  bullet & $J$=2--1 & $J$=3--2 & $J$=5--4 &  $J$=6--5 &  $J$=8--7 &  $J$=10--9  \\
  &    &  \multicolumn{6}{c}{$\int T_{MB}\,dV^{b}$}  \\
\hline
 ($-$30, +100)&   R1 & 4.7$\pm$0.2 & 4.7$\pm$0.4 & $<$0.5 & $<$0.7 & ... &  ... \\
\hline  
 ($-$30, +140) &   R &10.1$\pm$0.2 & 13.4$\pm$0.4 &7.8$\pm$0.9 & 2.0$\pm$1.2 & ... &... \\
 (-28.5, +137.5)  &    &  &  &  && 1.7$\pm$0.3 &...  \\
\hline
 (+22.5, $-$64.5)&   B1 & 19.3$\pm$0.3 & 28.9$\pm$0.4 & 24.9$\pm$0.9 & 18.8$\pm$1.2 &  &\\ 
 (+21.1, $-$62.5)   &      & &     &  &    & 8.7$\pm$0.4  & 2.6$\pm$0.6 \\
 (+21.1, $-$72.5)     &    &  &  &  && 4.6$\pm$0.4 &...  \\
\hline
 (+22.5, $-$36) &   B0 &7.4$\pm$0.3 & 9.5$\pm$0.4 & 6.2$\pm$0.7 & 4.4$\pm$1.2& ... & ... \\
 (+21.1, $-$42.5)    &    &  & & & & 2.5$\pm$0.3 & $<$0.8 \\
\hline
 (+35, $-$100) &   B2 & 12.7$\pm$0.3 & 14.4$\pm$0.4 & 6.5$\pm$0.7& 4.1$\pm$1.2 & & ... \\
 (+31.1, $-$102.5)  &   &  & &  & & 1.5$\pm$0.3& ... \\
 (+41.1, $-$102.5)  &    &  & &  & & 1.8$\pm$0.3 & $<1$ \\
\hline
 (+41.5,$-$110) &  B2  & 12.5$\pm$0.3 &15.2$\pm$0.4& 8.6$\pm$0.7&5.0$\pm$1.2 &&\\
 (+41.1, $-$112.5) &&&&  &  & 1.7$\pm$0.3 &  ...  \\	  
\hline\\[-5pt]
\end{tabular}

$^a$ Offset in arcsec with respect to L1157-mm, ($\alpha_{2000}$=20$^{h}$39$^{m}$06.2$^{s}$,
$\delta_{2000}$=68$^{d}$02$^{m}$15.9$^{s}$) \\
 $^b$ Flux errors refer to the rms noise in the individual spetra only. Total calibration uncertainty are of the order of 10\% for the $J\le$8
lines and 20\% for the $J$=10 and 11 lines.
\end{table*}

\section{Analysis of the SiO excitation}

For a quantitative analysis of the SiO excitation employing ratios 
among different lines, we need to correct the derived intensities 
for the different beams adopted for the observations. 
In L1448 this correction can be significant, especially for the B1 and R1 knots that 
are located at $~$20$\arcsec$ 
from the mm source. Each of these bullets is about 10\arcsec\ in length, 
but is not resolved in the transverse direction at a resolution of 
2$\arcsec$ (Guilloteau et al. 1992). Since we base our excitation analysis on 
the ratios between different lines, we have convolved all the observations along the
strip to the beam of the $J$=2--1 transition, in order to obtain corrected 
ratios on the B1 and R1 bullets. 
In L1157 the problem seems to be less severe, since SiO IRAM maps of different 
transitions show that the bullets emission region is large even for th e SiO 2--1 line taken 
with a 27$\arcsec$ beam (Bachiller at al. 2001). 
 In order to properly account for beam-filling effects, we have convolved the 8--7 data
to same resolution as the 2--1 beam. 
We find that the peak temperature of the line decreases by a factor from 1.3 in knot B2 to
 2.2 in knot R. These changes in the peak temperature are consistent with a size of
about 18-22 arcsec in knots R, B0, and B1 and around 30 arcsec for knot B2.
In the analysis we have considered both the case of uncorrected line ratios and line ratios
 corrected assuming the above estimated sizes in order to evaluate the effect of beam dilution on the
 final results.

\subsection{Large Velocity Gradient analysis}

\begin{figure}[!ht]
\resizebox{\hsize}{!}{\includegraphics{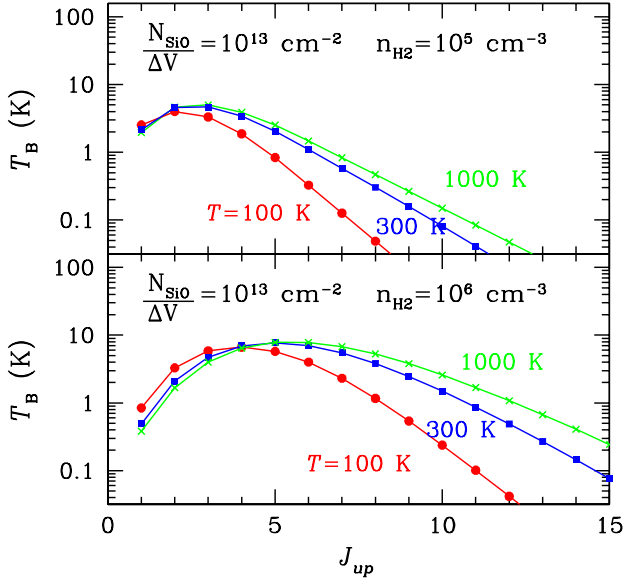}}
\caption{\label{fig:model}SiO brightness temperatures predicted by LVG model calculations for 
different temperatures and for densities of 10$^{5}$ cm$^{-3}$, (upper panel) and 10$^{6}$ cm$^{-3}$ (lower panel). 
A ratio between column density and velocity spread 
N$_{SiO}$/$\Delta V$ equal to 10$^{13}$\cmd\,km$^{-1}$\,s has been considered in this plot.}
\end{figure}

For the excitation analysis, we use an LVG model that considers the 
first 20 SiO levels, with collision rates from Turner et al. (1992)
up to 300 K, plus  the extrapolations given by Sch\"oier et al. (2005) 
for larger temperatures (up to 2000 K).  
In Fig. \ref{fig:model} we plot the expected line brightness as a 
function of the rotational quantum number, for different values of 
the gas kinetic temperature and H$_2$ density. 
While lines with $J_{up} <$ 5 do not strongly depend on the 
temperature but are sensitive to the density, high temperatures 
regimes (T$>$100 K) can be distinguished from low temperature gas only 
through high-J ($J\ge 5$) lines. The plots also show that the lines 
become insensitive to the temperature for T$\ga$500 K.
These plots have been obtained using a ratio between column density and velocity 
dispersion (N$_{SiO}$/$\Delta 
V$) of  10$^{13}$ \cmd\,km$^{-1}$\,s. In this specific case, the maximum line 
opacity is $\sim$ 0.2. Opacity effects start to become significant, 
affecting the line ratios up to the point at which the lines become
insensitive to the SiO abundance, for N$_{SiO}$/$\Delta 
V \sim$ 5$\times 10^{13}$\cmd\,km$^{-1}$\,s, i.e. for column densities of the order of 
5$\times 10^{14}$\cmd\, in gas with total velocity about 10\,\kms. Since this value is significantly 
larger than the typical SiO column densities derived in shocked regions (i.e. Codella et al. 1999),
we have assumed in the first instance that the lines are optically thin in order to 
derive 
temperature and density from the different line ratios. Large values of 
N$_{SiO}$/$\Delta V$ have been introduced only when a 
reasonable fit could not be obtained assuming optically-thin emission. 
Once $T$ and $n_{H_2}$ have been estimated, the SiO column density
has been derived from the absolute intensity of the $J$=5--4 line.
We selected this line because is the one with the smaller beam having
good S/N ratios. From the derived column density and the velocity 
dispersion measured on our spectra, we have checked a posteriori 
the validity of our assumption of low opacity.

We have applied the LVG code on the intensities of each of the observed bullet,
integrated in all the velocity channels, separating only the 
blue-shifted and red-shifted components.
The results of the LVG analysis are discussed
below for L1448 and L1157. 

\subsubsection{L1448}

\begin{figure}[!ht]
\resizebox{\hsize}{!}{\includegraphics{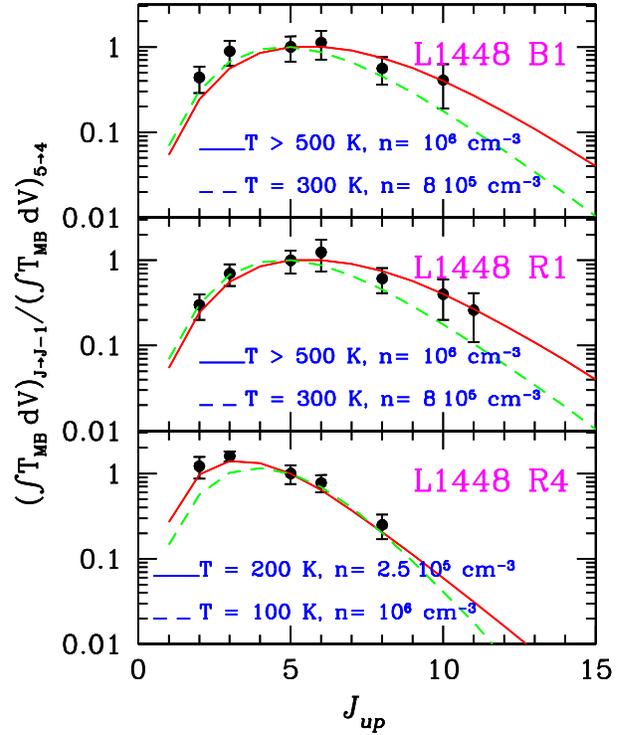}}
\caption{\label{fig:1448fit}
SiO line intensities observed in knots B1, R1 and R4 of the L1448
outflow, normalized to the $J$=5--4 intensity and compared
with the LVG model computations. Solid lines refer to the best fit
through the data, while the dashed line represents the model fit at the
lowest temperature which is still compatible with the data errors.
The plotted line ratios uncertainty consider both the individual rms errors and an absolute calibration
uncertainty of 10\% for the lines with $J\le$8 and 20\% for the $J$=10 and 11 lines}
\end{figure}

Fig. \ref{fig:1448fit} shows the observed integrated intensities, 
normalised to the 5--4 transitions, of the bullets sampled by our observations,
compared with the LVG model computations. Errors on line ratios take into account both the rms in the 
spectra and the
calibration uncertainties.
Assuming a velocity dispersion of $~$20\kms, 
i.e. comparable to the FWHM of the observed lines, the modelled line ratios are insensitive to
the column density up to values $~$10$^{15}$\cmd.
 Since our observations do not fully sample the region covered
by the larger 2--1 beam, the applied convolution may not properly takes into 
account all the emission region. In addition to this, 
the lack of a good fit may signal the presence of gradients in the physical parameters, 
with components at lower density/temperature encompassed by the large beams.
Consequently, we adopted the
best model that fits only the lines with $J \ge$5, which have been 
obtained through a more similar beam size and thus trace similar gas components.
 A density of $n_{H_{2}}$ = 10$^6$ \cmt\,
is well constrained by the observed ratios, while any temperature $T\ga$500 K
gives an equally good fit.  We have also considered models assuming a 
higher density and lower temperature to explore whether other pairs
of values can fit the data equally well. We find that within the errors,
the line ratios of knots B1 and R1 can  be consistent with a 
temperature of 300 K (see Fig. 5). 

In the bullet R4 the best fit is obtained for $T$=200 K and 
$n_{H_{2}}$=2.5$\times 10^5$\cmt\, consistent with the lower excitation 
nature of this clump with respect to B1 and R1. 

SiO column densities, determined from the absolute observed 
intensity of the 5--4 line, are of the order of 3--3.5$\times 10^{13}$ 
\cmd\, in all the three bullets. Values derived for the B1 and R1 
clumps, however, are probably just lower limits, since the emission 
region does not fill the entire 11 arcsec beam of the 5--4 observation. 
If the emission region is confined to a jet-like knot  about 
2$\arcsec$ in diameter (see e.g. Guilloteau 1992 ), the corresponding column density would be 
enhanced by about a factor of 4, thus being of the order of 
10$^{14}$ \cmd. 

\subsubsection{L1157}

\begin{figure}[!ht]
\resizebox{\hsize}{!}{\includegraphics{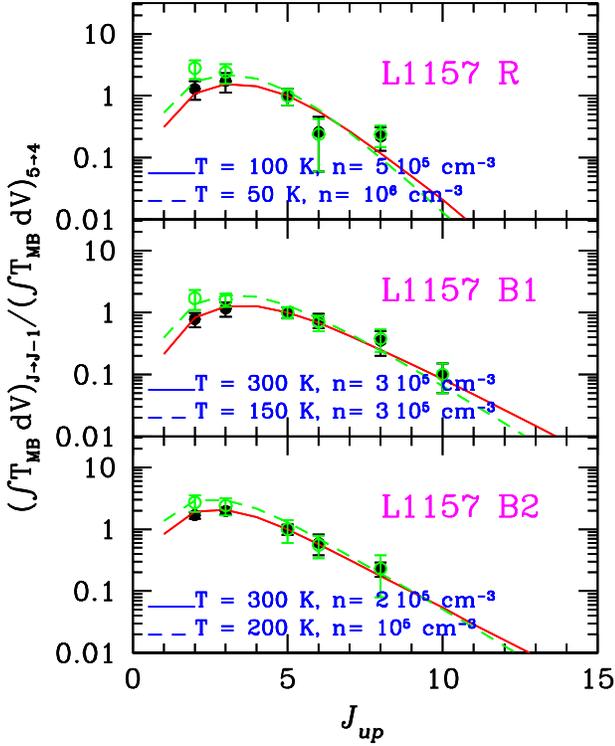}}
\caption{\label{fig:1157fit} The same as Fig. \ref{fig:8su5} but for the knots R, B1 and B2 of L1157.
 Filled circles refer to observed ratios assuming beam filling equal 1, while open circles are the
observed ratios corrected assuming an emitting size of $\sim$20\arcsec for R and B1, and 30\arcsec for B2
(see Section 3). The best fits obtained considering only the $J\le$5 lines (solid) and all the beam
filling corrected lines (dashed) are reported.}
\end{figure}

 Fig. \ref{fig:1157fit} shows the observed line intensities, normalized to the $J$=5--4 transition,
 in the bullets R, B1 and B2, where we performed the LVG analysis. In the figure, filled 
 circles represent line ratios not corrected for beam filling effects; open circles 
 represent ratios corrected for beam filling, assuming the emission sizes estimated by convolving the $J$=8--7
 observations to the $J$=2--1 beam (see previous section). This correction significantly affects only the 
 2--1 and 3--1 transitions, so we firstly fitted only the lines with $J \ge$5.
In B1 and B2 the line ratios are consistent with a temperature 
of $\sim$ 300 K, a density $\sim$3$\times 10^5$ \cmt\, and optically-thin lines. The bullet R, on the other hand,  
is consistent with colder gas at $T \sim$100 K. These fits reproduce rather well the 2--1 and 3--2 
uncorrected ratios. We have also tried to reproduce all the beam-filling-corrected data points, obtaining
the fit indicated by dashed lines in Fig. \ref{fig:1157fit}. Good fits are obtained also in this case, by
adopting slighter lower temperatures of $\sim$50--200 K.
Column densities are 2, 5 and 8$\times 10^{13}$cm$^{-2}$ at bullets R, B2 and B1 respectively. Therefore, the L1157 clumps appear less excited 
than the L1448 B1 and R1 bullets, confirming a dependence of the excitation 
conditions with the velocity of the bullets.

\subsection{Excitation as a function of velocity}

The above LVG analysis has been performed on intensities 
integrated over all the observed velocity channels, thus giving
physical parameters averaged over all the emitting gas. 
The observed line profiles, however, are not symmetric and clearly show that gas components
at different velocities are seen along the line of sight.
We can then investigate if these components are associated with different excitation conditions.
 
\begin{figure}[!ht]
\resizebox{\hsize}{!}{\includegraphics[angle=-90,scale=0.8]{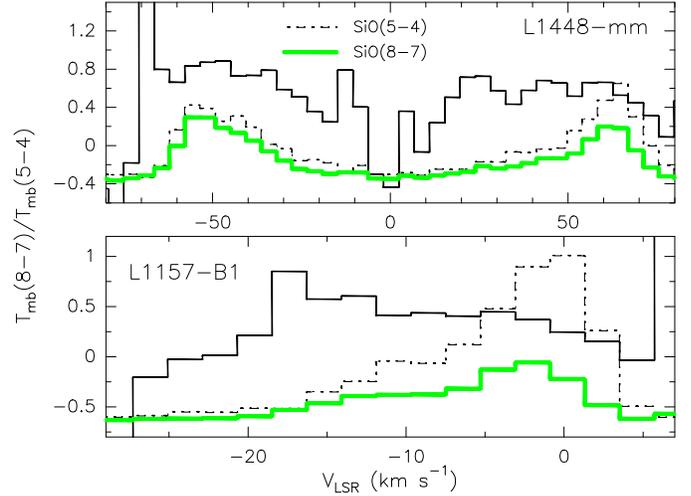}}
\caption{\label{fig:8su5} Comparison between the intensity profiles vs velocity of the 
SiO 5--4 and 8--7 lines in the L1448 R1 and B1 bullets (upper) and  the L1157 B1 bullet (lower). 
The dot-dashed and grey lines are the individual profiles of the 5--4 and 8--7 lines, respectively,
while the black line represents the profile of the line ratio. There is a trend, more evident for 
L1157 B1 and L1448 B1, showing that this ratio increases with the velocity. }
\end{figure}

Fig. \ref{fig:1157-spec} shows that in the L1157-B1 spectra there are at least 
two components, one peaking close to ambient velocity and one at about $-$15\kms, 
whose relative $T_B$ ratio increases going from the 2--1 to the
 8--7 line (see also Fig. 8). The signal-to-noise ratio in the line wings of the high-J lines is 
not high enough to perform a separate LVG analysis in the different 
velocity channels. We can however still infer a quantitative trend in the 
excitation as a function of velocity by comparing sensitive line ratios. In Fig. \ref{fig:8su5}
we have plotted the SiO 8--7/5--4 ratio as a function of velocity for
the L1157 B1 bullet and for the L1448 R1 and B1 bullets. The 8--7 and 
5--4 lines have been observed with similar beamsizes and should thus be
not much affected by different beam fillings. 

For L1157-B1 the 8--7/5--4 ratio is $\sim$0.25$\pm$0.03 in the peak close to ambient V$_{LSR}$ velocity 
and $\sim$0.6$\pm$0.25 at velocities $\sim -$15\kms.
In L1448 B1/R1, line ratio differences with velocity 
are less clear, although a trend is still recognisable in the B1 bullet, where
the 8--7/5--4 ratio varies from  $\sim$0.2-0.3$\pm$0.2 at V$_{LSR} \sim -$20\kms\,and 
$\sim$0.8$\pm$0.08 at V$_{LSR} \sim -$50\kms. 
A ratio $\sim$0.2-0.3 is consistent with gas at
a density of $\sim$10$^6$ \cmt\, and temperature $\sim$200 K, while a 
value of 0.5-0.7 would require either a larger temperature ($>$ 500 K) or
a larger density ($\sim$5$\times 10^6$ \cmt). 

In L1448 B1/R1 the observed excitation as a function of velocity is consistent 
with gas coming from 
different layers behind a shock:  the maximum
temperature is found at the shock front, where the gas is compressed and the 
velocity is close to the shock velocity, and lower temperatures occur further out where the gas has 
been slowed down. In L1157-B1, however, we find a puzzling result, since most of the 
gas emission comes from the lower
excitation component close to the ambient velocity, a feature that is not predicted by shock models, 
as we will further discuss in Sect. 4.  
Jim\'enez-Serra et al. (2004) have suggested that SiO emission at ambient velocity associated 
with outflows could be a signature of the interaction between the magnetic shock precursor 
in a C-type shock, with the ambient pre-shock gas. This interpretation in the case of L1157-B1 is however 
in contrast with the large SiO abundance of the low velocity component.
In fact, the derived velocity averaged column density of 8$\times 10^{13}$\cmd\, is almost
entirely dominated by the gas component at ambient velocity. Such a high column density
implies that this component should have experienced significant shock reprocessing.
In addition, the ambient component due to the precursor is expected to have very narrow widths,
of the order of 1\kms\,or less, as observed by Jim\'enez-Serra et al. (2004) in L1448, while
in L1157 the width of the component at ambient velocity is $\sim$7\kms.
The inferred different excitation conditions of the low and high-velocity components 
in L1157 could be instead due to an evolutionary effect, where the SiO gas at ambient velocity in 
L1157 arises from decelerated post-shocked material, that has already slowed down, as suggested
in Codella et al. (1999).

The velocity structure of the L1448 B1 and R1 bullets,
coupled with the inferred variations of physical parameters with the velocity, 
resemble very closely those found in optical/IR jet-knots near the 
protostar. Indeed, the base of the optical jets structure 
is characterised by the presence of both
low and high velocity components, with this latter being the brightest and 
having the highest excitation conditions (e.g. Ray et al. 2006). 
This similarity, and the evidence that the collimated EHV bullets are hotter and denser than 
usually found in the CO entrained outflow, suggest that 
they may indeed trace the main jet from the embedded protostar. Such an interpretation has 
been also suggested for the EHV SiO jet of HH211 on the basis of its excitation and velocity 
structure derived from SMA observations (Hirano et al. 2006).

\begin{table}
\caption{Physical parameters of the observed bullets}
\begin{tabular}{lccccc}
\hline
& \multicolumn{2}{c}{SiO} & \multicolumn{2}{c}{CO($J>$14)$^{a}$}
& \multicolumn{1}{c}{H$_2$ 0--0$^{b}$}\\
& $T_{\rm kin}$ & $n_{H_2}$ & $T_{\rm kin}$ & $n_{H_2}$ & $T_{\rm kin}$ \\
& K & cm$^{-3}$ & K & cm$^{-3}$ & K \\
\hline
\multicolumn{6}{c}{L1448}\\
\hline
{B1} & $>$500 & 8$\times 10^{5}$ & 1200 & 6$\times 10^{4}$ & 950--1900 \\
{R1} & $>$500 & 10$^{6}$ & 1200 & 6$\times 10^{4}$ & 950--1900 \\
{R4} & 200 & 2.5$\times 10^{5}$ & 600--650 & 1-3$\times 10^{4}$ &600--650 \\ 
\hline	
\multicolumn{6}{c}{L1157}\\
\hline
{B1} & 150--300 & 3$\times 10^{5}$ &  350--800 & 5-60$\times 10^{5}$ & 560--1040\\
{B2} & 200--300 & 2$\times 10^{5}$ & 350--800 & 5--60$\times 10^{5}$ & 560--1040 \\
{R} & 50--100 & 1--5$\times 10^{5}$ & ... & ...& 570--860\\
\hline\\[-5pt]
\end{tabular}

~$^a$ Derived from ISO-LWS data: Nisini et al. (2000) for L1448 and Giannini
et al. (2001) for L1157.\\
~$^b$ Derived from ISO-SWS and ISO-CAM data: Nisini et al. (2000) for
L1448 and Cabrit et al. (1998) for L1157.
\end{table}

\section{Comparison with other warm gas tracers}

In the physical conditions traced by the SiO emission, i.e.
$T\sim$100--500 K and $n\sim$10$^{5}$-10$^{6}$ cm$^{-3}$, the main 
gas coolants are pure-rotational H$_2$ lines,
CO transitions with $J_{up}$ between 10 and 20, and H$_2$O 
rotational lines. H$_2$O cooling is only significant
if the conditions are appropriate to have 
large water abundances; this is indeed the case in low-velocity
C-type shocks where most of the gas-phase SiO is expected to be 
produced. Most of the cooling from these molecules occurs at mid
and far IR wavelengths. Since both the investigated outflows have been 
observed with ISO we can compare our 
diagnostic on the SiO emission with the results obtained by ISO on the 
other main coolants (see Table 4). 
In L1448, strong CO with $J>$14 and H$_2$O emission 
is associated with the B1 and R1 bullets and the inferred physical 
conditions are $T$=1200 K and $n_{H_2}$=6$\times 10^{4}$ cm$^{-3}$ (Nisini et al. 1999, 2000). The 
 temperature is also consistent with the H$_2$ 0--0 lines detected by SWS.
The SiO emission seems to be confined to a denser region 
with respect to the other coolant, which implies a large density 
stratification inside the bullets. In 
the R4 bullet the conditions derived from the ISO-observed CO lines 
are $n_{H_2} \sim$1--3$\times 10^{4}$ cm$^{-3}$ and $T \sim$600\,K.
These are consistent with the lower excitation 
found in this position with SiO, although for this latter a higher density and lower
temperature are estimated. From the comparison of our derived SiO column 
densities ad the N(CO) estimated from the ISO data ($\sim$10$^{17}$ cm$^{-2}$, Nisini et al. 2000), we obtain a SiO abundance 
of $\sim$10$^{-7}$ and 3$\times 10^{-8}$ in the B1/R1 and R4 positions, respectively
(Table 5).

In L1157 the temperature and density estimated from the ISO data ranges
between 350--800 K and 5$\times 10^{5}$--6$\times 10^{6}$ cm$^{-3}$ respectively
(Giannini et al. 2001), thus fairly consistent with the parameters derived 
from SiO. Temperatures in the same range are also found from the H$_2$ 0--0
lines observed with ISOCAM (Cabrit et al. 1998).
Therefore it appears that the physical conditions in the B1 bullets are more
uniform, without strong gradients. The SiO abundance is, in both B1 and R bullets, 
of the order of 6--8$\times 10^{-8}$.

From the comparison of the SiO column density and the column density of the warm water traced by ISO, we derive a SiO/H$_2$O abundance ratio of 
$\sim$ 2--3$\times 10^{-4}$ in L1448 and 
$\sim$10$^{-3}$ in L1157 (see Table 5). The difference in these two values 
cannot be explained by the different excitation conditions of the
two flows.
For low velocity shocks, the SiO/H$_2$O ratio is sensitive to the shock speed, since shocks with velocity $\le$ 20\kms 
do not efficiently release  
silicon by sputtering of dust grains, while still being very efficient to produce of large H$_2$O 
abundances by means of high temperature (T$\ga$300 K) chemical reactions. 
However, both the maximum velocity and the excitation 
conditions observed in L1157 indicate that here the shock speed should be lower than in L1448.
The different SiO/H$_2$O ratio observed could be due instead to an evolutionary effect, if the two molecules have different destruction timescales after the passage of the shock. 
The timescales for H$_2$O depletion onto the dust grains range between 10$^4$--10$^5$
 yr depending on the density (Bergin et al. 1998). Similar timescales are also predicted 
for the SiO destruction either by depletion on dust grains or by reactions with OH which bring 
to the formation of SiO$_2$ (Codella et al. 1999, Schilke et al. 1998). 
In conclusion, no simple considerations based on excitation conditions and evolutionary timescales
seem to explain the different observed abundance ratio. More detailed chemical models of the
two regions are required for a proper understanding of this point.


\begin{table}
\caption{SiO and H$_2$O abundances}
\begin{tabular}{lccc}
\hline\\
& N(SiO)(cm$^{-3}$) & X(SiO) &X(SiO)/X(H$_2$O)$^a$\\
\\
\hline
{L1448 B1--R1} & $\sim$10$^{14}$ & $\sim$10$^{-7}$ & $\sim$2$\times 10^{-4}$\\
{L1448 R4} & 3$\times 10^{13}$ & 3$\times 10^{-8}$ & 3$\times 10^{-4}$\\ 
\hline
{L1157 B1} & 8$\times 10^{13}$ & 8$\times 10^{-8}$ & 10$^{-3}$\\
{L1157 R} & 2$\times 10^{13}$ & 6$\times 10^{-8}$ & $\sim$10$^{-3}$\\
\hline\\
\end{tabular}
\\
~~~$^a$ Assuming the H$_2$O abundance derived by Nisini et al. (2000)(L1448)
and Giannini et al. (2001)(L1157).
\end{table}

\section{Comparison with predictions from shock models}

The multi-line analysis performed can be used to check if the derived excitation
conditions and abundances are in agreement with the expectations from
shock models. 

The SiO molecule is produced in C-type shocks by the injection
into the gas-phase of silicon by sputtering and/or
grain-grain collisions, followed by gas-phase reactions with O and O$_2$
(Schilke et al. \cite{schilke}, Caselli et al. \cite{caselli}). 
To be efficient in producing the large
observed column densities of SiO by sputtering, such shocks need to have speeds of at least 
25 \kms. At lower velocities (V$_s \sim$ 20\kms) and high density, grain-grain collisions 
may be more efficient than sputtering for the production of gas-phase 
elemental silicon. At such speeds, the neutral temperature of 
the post-shocked gas rises above 1000 K and then gently declines to return
to its pre-shock values in a region of $\sim$5$\times 10^{16}$ cm ($\sim$10\arcsec at a distance 
of 350 pc, i.e. comparable with the bullets dimensions). 

We can compare the observed line intensity distributions as a function 
of $J$ (Figs. 5,6), with those predicted by the Schilke et al. (1987) shock models 
as a function of 
pre-shock density and velocity (see their Fig. 6). The line ratios 
derived for the L1448 bullets are consistent with n(H$_2$) $\sim$ 10$^{6}$\,cm$^{-3}$ 
and shock velocities larger than 30 \kms. L1157 B1 is instead 
better fit with a shock at lower density, between 10$^{5}$ and 10$^{6}$ cm$^{-3}$.
The models are not very sensitive to the shock velocity for v$_s$ larger 
than 30 \kms. This is a consequence of the fact that the 
post-shocked temperature is a function of the velocity, and that the 
SiO line ratios become insensitive to temperature variations for $T$
larger than $\sim$500 K. 
However, the peak post-shock temperature reaches values larger than 1000 K 
already in a 25\,\kms\ shock (e.g. Kaufman \& Neufeld  \cite{kauneu}), 
thus the SiO emission region is indeed expected to be very warm.
Observations of submillimetre tracers with higher excitation energies, such as
the high-J CO lines, are needed to better constrain the temperature values inside the bullets.

The fractional SiO abundances derived in Table 5 are also consistent with 
those predicted by models in the regions of the post-shocked gas
having the larger temperatures (between 10$^{-8}$ and 10$^{-7}$). In further out 
regions, where the temperature 
drops close to the pre-shock values, SiO is in part converted to SiO$_2$ 
by reactions with OH, and its abundance decreases to values below 
10$^{-9}$. The averaged high observed abundances therefore testify that we are 
indeed observing the warmer regions just behind the shock front.

We can finally compare the observed line profiles 
with that predicted by models. Schilke et al. (1997) computed 
how the profiles change 
as a function of the rotational quantum number, both in the case of  
sputtering of Si-bearing material in grain cores and for the case
of mantle sputtering. The  overall profiles produced in the two cases 
are similar to those observed in the L1448 bullets: very asymmetric, 
with a gradual decrease of intensity for velocities lower than peak velocity 
and a very sharp decrease on the other side. 
In these bullets the intensity peaks occur at a V$_{LSR}$ velocity of about 60\kms\,which, 
deconvolved for inclination effects,
corresponds to a total velocity larger than 150 \kms\, (for an inclination angle of 69$^{\circ}$, Girard \& Acord \cite{ga}). 
This is much larger than the estimated shock velocity, implying that probably the bullet is moving 
in a medium that has been already put into motion, and thus that the shock occurs 
at the much lower velocity given by the differential speed among the 
two fluids. 

In contrast with L1448, the L1157-B1 profiles do not appears consistent with the predicted ones, 
since the brightness peak is close to the ambient velocity and the intensity declines 
towards the higher velocity. We have already pointed out and discussed this behaviour in the previous sections, suggesting that 
the peak at ambient velocity could be due to a shock that has already slowed down but still retaining the original SiO abundance. 
Models taking into accounts time evolution effects are needed to test this hypothesis. In addition, geometrical effects can be also 
important and are expected to play a role in the line profile appearance.
Bow shock models, for example, produce line profiles that are highly dependent upon the
bow inclination angle and for some configurations produce shapes similar to those observed in 
the L1157 SiO line (see e.g. Schultz et al. 2005).

Another effect predicted by the Schilke et al. (1997) model is that the 
 line profile broadens for lines with increasing $J$.
Such a trend is not seen in our observed lines. 
In Fig. 8, for example, the profiles of the 2--1 and 8--7 lines are compared: 
in both the objects (L1448 and L1157) the two lines are equally broaden.

In conclusion, current shock models fit reasonably well 
 the average  physical conditions and abundances of the SiO gas derived from 
 our multiline analysis in both the EHV bullets of L1448 and in the lower velocity clumps of L1157. 
 However, the line profiles are still not well reproduced,
especially when profile variations among lines with different excitation are taken into account. 
Effects such as a time-dependent chemistry and a geometry different from plane-parallel need
 to be taken into account for a more quantitative comparison with the observed profiles.
In addition, Interferometric observations should be able to
resolve spatially the emitting regions contributing to the different components, thus allowing 
us to constrain the geometry to be adopted in a more detailed modelling. It would be also instructive 
to compare the observed SiO profiles with the profiles of other molecules similarily affected by
a time dependent chemistry. H$_2$O, in particular, will be studied in details by means of the Herschel
observatory, providing valuable constraints on the shock model chemistry.

\begin{figure}[!ht]
\resizebox{\hsize}{!}{\includegraphics[angle=-90,scale=0.8]{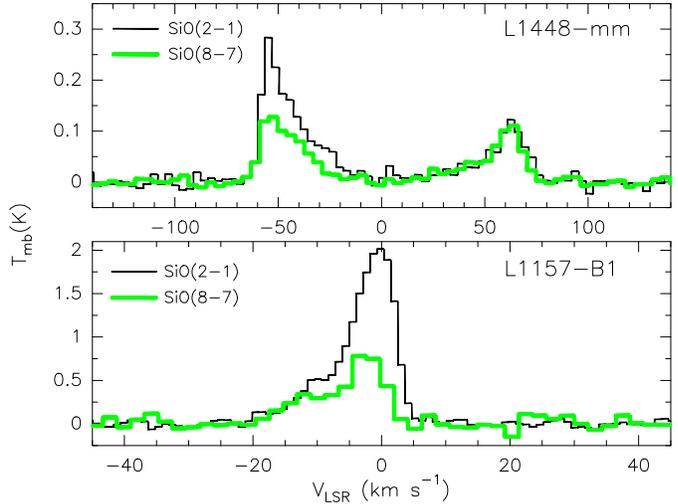}}
\caption{\label{fig:8con2} Comparison between the intensity profiles of the 
SiO 2--1 (black) and 8--7 (gray) lines in the L1448 R1 and B1 bullets (upper) and  the 
L1157 B1 bullet (lower). The L1448 8--7 line profile toward the mm source central position has been
convolved to the beam of the 2--1 transition in order to compare the same emitting area.}
\end{figure}

\section{Conclusions}
In this paper we present the first SiO multi-line analysis (from $J$=2--1 to $J$=11--10) 
of the molecular bullets along the outflows of the Class 0 sources L1448-mm and L1157-mm. 
In L1448-mm, the  bullets represent collimated and regularly spaced clumps travelling at velocities of
hundreds of \kms. In contrast, the L1157 bullets have lower velocities and are 
characterised by a rich chemistry. From our analysis, we investigated the physical properties of these different types
of bullets, exploring their excitation conditions and the 
capability of current models to reproduce the various SiO line ratios and profiles. The main results of our analysis are the following:
\begin{itemize}
\item In L1448, the SiO line ratios of the EHV molecular bullets B1 and R1 close to the driving source are consistent with a gas at $n_{H_2} \sim$10$^6$ cm$^{-3}$
and $T_{kin} \ga$ 500 K. Conversely, the bullet R4, located at the apex of the red lobe molecular outflow and very strong in the SiO 2--1 emission, is associated with lower excitation gas 
($n_{H_2}$ = 2.5\,10$^5$ cm$^{-3}$ and $T_{kin}=$\,200\,K).
Also the low velocity bullets in L1157 show lower excitation conditions, with $n_{H_2} \sim$1--5$\times 10^{^5}$ cm$^{-3}$
and $T_{kin} \sim$ 100--300 K. In both the sources, there is however a clear evidence of the presence of 
velocity components having different excitation conditions, with the denser and/or warmer gas associated with 
the gas at the higher speed.

\item While in L1448 the bullets emission is dominated by the highly excited 
and high velocity gas, in L1157 the peak of the emission at the ambient velocity is associated 
with the lower excitation conditions. Such a difference could be due to evolutionary effects, 
where the L1157 SiO gas at ambient velocity is due to an older shock that has already 
slowed down and is progressively reaching the physical conditions of the ambient medium.

\item The physical conditions averaged over all the velocity components are consistent with those inferred by means of mid- and far-IR lines observed with the ISO satellite. In L1448, however, the high-J CO lines detected by ISO-LWS trace a gas with a density about an order of magnitude lower than that measured with the SiO lines, indicating that gradients in the physical parameters can be present in the bullets region. 

\item The column density of the SiO gas does not show large variations in the different bullets, ranging between 2$\times 10^{13}$ and 10$^{14}$ cm$^{-2}$. 
A comparison with the ISO-LWS derived abundance of H$_2$O , a molecule
also very sensible to the shock conditions, shows that the SiO/H$_2$O abundance ratio is $\sim$ 2--3$\times 10^{-4}$ in L1448 and 
$\sim$10$^{-3}$ in L1157. Such a different ratio is not easily explained by either the local excitation conditions and the different flow evolutionary timescales.

\item Finally, we find that current shock models for the SiO excitation reproduce quite well the observed line ratios when shocks with velocities larger than $\sim$ 25\kms\, are considered. The models however fail to predict the observed profile variations as a function of the line excitation, as well as the 
strong emission at ambient velocity observed in L1157. 
Effects such as a time-dependent chemistry and a geometry different from plane-parallel need
 to be taken into account for a more quantitative comparison with the observed profiles.

\item We also point out that the velocity structure 
   and the variations of physical parameters with the velocity observed in the EHV bullets in L1448
   resemble very closely those found in optical/IR jet-knots near the 
   protostar, suggesting that the mm collimated bullets have a similar origin and excitation 
   mechanism. 
\end{itemize}

\begin{acknowledgements}
B.N., T.G. and C.C.  wish to acknowledge support through the Marie Curie Research Training Network 
JETSET (Jet Simulations, Experiments and Theory) under contract MRTN-CT-2004-005592.
The James Clerk Maxwell Telescope is operated by The Joint Astronomy
Centre on behalf of the Particle Physics and Astronomy Research Council
of the United Kingdom, the Netherlands Organisation for Scientific
Research, and the National Research Council of Canada.
\end{acknowledgements}


\begin{thebibliography}{}

\bibitem[2001]{arce1}
Arce H.G., Goodman A.A., 2001, ApJ 554, 132
\bibitem[2002]{arce2}
Arce H.G., Goodman A.A., 2002, ApJ 575, 928 
\bibitem[2001]{bach01}
Bachiller R., Per\'ez Guti\'errez M., Kumar M.S.N., Tafalla M., 2001, A\&A 372, 899
\bibitem[1990]{bach90}
Bachiller R., Mart\'{\i}n-Pintado J., Tafalla M., Cernicharo J., Lazareff B., 1990, A\&A 231, 174
\bibitem[1998]{bergin}
Bergin, E. A., Melnick, G. J., \& Neufeld, D. A. 1998, ApJ, 499, 777  
\bibitem[1997]{bourke}
Bourke T.L., Garay G., Lehtinen K.K., et al., 1997, ApJ 476, 781
\bibitem[1998]{cabrit}
Cabrit, S., Bontemps, S., Lagage, P.O. et al. 1998 in {\it The Universe as seen by ISO}, p.449,
eds. P. Cox \& M.F. Kessler
\bibitem[1997]{caselli}
Caselli, P., Hartquist \& T.W., Havnes, O. 1997, A\&A, 322, 296
\bibitem[2001]{chandler}
Chandler C.J., Richer J.D., 2001, ApJ 555, 139
\bibitem[1999]{cod99}
Codella C., Bachiller R., Reipurth B., 1999, A\&A 343, 585
\bibitem[1997]{dutrey}
Dutrey A., Guilloteau S., Bachiller R., 1997, A\&A 317, L55
\bibitem[2001]{gia}
Giannini T., Nisini B., Lorenzetti D., 2001, ApJ 555, 40
\bibitem[2001]{ga}
Girard J.M., Acord, J.M.P., 2001, ApJ 552, L63
\bibitem[2004]{gibb}
Gibb A.G., Richer J.S., Chandler C.J., Davis C.J., 2004, ApJ 603, 198
\bibitem[1999]{guethgui}
Gueth F., Guilloteau S., 1999, A\&A 343, 571 
\bibitem[1992]{guill}
Guilloteau S. Bachiller, R, Fuente, A., Lucas, R., 1992, A\&A, 265, L49
\bibitem[1997]{hatchell}
Hatchell J., Fuller G.A., Ladd E.F., 1999, A\&A 346, 278
\bibitem[2006]{hirano}
Hirano, N., Liu, S., Shang, H. et al., 2006, ApJL 636, L141
\bibitem[2004]{jimenez}
Jim\'enez-Serra, I., Mart\'in-Pintado, J., Rodr\'iguez-Franco, A., Marcelino, N., 2004, 
ApJ, 603, L49
\bibitem[1996]{kauneu}
Kaufman, M.J., Neufeld, D.A., 1996, ApJ, 456, 61
\bibitem[1992]{martin}
Mart\'{\i}n-Pintado J., Bachiller R., Fuente A., 1992, A\&A 254, 315
\bibitem[1999]{nisini99}
Nisini, B.; Benedettini, M.; Giannini, T. et al., 1999, A\&A 350 529
\bibitem[2002]{nisini00}
Nisini B., Codella C., Giannini T., Richer J.S., 2002, A\&A 395, L25
\bibitem[2002]{nisini02}
Nisini B., Benedettini M., Giannini T., Codella C., Lorenzetti D., Di Giorgio A., 
Richer J.S., 2000, A\&A 360, 297
\bibitem[2006]{palau}
Palau A., Ho P.T.P., Zhang Q., et al. 2006, A\&A 636, L137
\bibitem[2000]{richer}
Richer J.S., Shepherd D.S., Cabrit S., Bachiller R., Churchwell E., 2000,
in Mannings V., Boss A.P., Russel S.S., eds, Protostars and Planets IV, Univ. Arizona Press,
Tucson, p. 867
\bibitem[1997]{schilke} 
Schilke, P., Walmsley, C.M. \& Pineau des For\^{e}st, G., Flower, D.R. 1997, A\&A, 321, 293
\bibitem[2005]{sch}
Sch\"oier, F. L. van der Tak, F. F. S. van Dishoeck, E. F. Black, J. H. 2005, A\&A, 432, 369
\bibitem[2005]{schultz}
Schultz, A. S. B., Burton, M. G., Brand, P. W. J. L. 2005, MNRAS, 358, 1195-1214
\bibitem[2004]{tafalla}
Tafalla M., Santiago J., Johnstone D., Bachiller R., 2004, A\&A 423, L21
\bibitem[1992]{turner}
Turner B.E., Chan K.W., Green S., Lubowich D.A., 1992, ApJ, 399, 114
\bibitem[1998]{van}
van Dishoeck E.F., Blake G.A., 1998, ARA\&A 36, 317
\bibitem[2004]{viti}
Viti S., Codella C., Benedettini M., Bachiller R., 2004, MNRAS 350, 1029

\end{thebibliography}
\end{document}